\documentclass[12pt]{spie} 
\usepackage{graphicx}
\usepackage{amssymb} 
\usepackage{amsmath}    
\usepackage{latexsym}   
\usepackage{bm}
\usepackage{times}

\def\abr{\mathrel{\stackrel{\longrightarrow}{A\!B}}\!}
\def\bar{\mathrel{\stackrel{\longrightarrow}{B\!A}}\!}

\newtheorem{prop}{Proposition}

\title{Logical synchronization: how evidence and hypotheses \\ steer
  atomic clocks}

\author{John M. Myers\supit{a} and F. Hadi Madjid\supit{b}
\skiplinehalf
\supit{a}Harvard School of Engineering and Applied 
Sciences, Cambridge, MA 02138, USA; \\
\supit{b}82 Powers Road, Concord, MA 01742, USA}

\authorinfo{Further author information: (Send correspondence to J.M.M.)\\J.M.M.: 
E-mail: myers@seas.harvard.edu, Telephone: 1 617 495 5263\\  
F.H.M.: E-mail: gmadjid@aol.com, Telephone: 1 978 369 1808}

\begin{document}
\maketitle

\begin{abstract}
A clock steps a computer through a cycle of phases.  For the propagation
of logical symbols from one computer to another, each computer must
mesh its phases with arrivals of symbols from other computers.
Even the best atomic clocks drift unforeseeably in frequency and phase; feedback
steers them toward aiming points that depend on a chosen wave function
and on hypotheses about signal propagation.  A wave function, always
under-determined by evidence, requires a guess.  Guessed wave
functions are coded into computers that steer atomic clocks in
frequency and position---clocks that step computers through their
phases of computations, as well as clocks, some on space vehicles,
that supply evidence of the propagation of signals.
Recognizing the dependence of the phasing of symbol arrivals on
guesses about signal propagation elevates `logical
synchronization.'  from its practice in computer engineering to a
dicipline essential to physics.  Within this discipline we begin
to explore questions invisible under any concept of time that
fails to acknowledge the unforeseeable.  In particular, variation
of spacetime curvature is shown to limit the bit rate of logical
communication.
\end{abstract} 

\section{Introduction}
As is well known, the atom or atoms in the atomic clock are passive---they
do not ``tick''---so the clock needs an active oscillator in addition to
the atom(s).  In designing an atomic clock to realize the second as a
measurement unit in the International System of Units (SI) , one encounters
two problems: (a) The resonance exhibited by the atom or atoms of the clock
varies with the details of the clock's construction and the circumstances
of its operation; in particular the resonance shifts depending on the
intensity of the radiation of the atoms by the oscillator.  (b) The
oscillator, controlled by, in effect, a knob, drifts in relation to the
knob setting.  Problem (a) is dealt with by introducing a wave function
parametrized by radiation intensity and whatever other factors one deems
relevant.  The SI second is then ``defined'' by the resonance that "would
be found" at absolute zero temperature (implying zero radiation).  For a
clock using cesium 133 atoms, this imagined resonance is declared by the
General Conference of Weights and Measures to be 9 192 631 770 Hz, so that
the SI second is that number of cycles of the radiation at that imagined
resonance \cite{sp330}.  To express the relation between a measured
resonance and the imagined resonance at 0 K, a wave function is chosen.
Problem (b) is dealt with by computer-mediated feedback that turns the knob
of the oscillator in response to detections of scattering of the
oscillator's radiation by the atom(s) of the clock, steering the oscillator
toward an aiming point.

A key point for this paper is that the wave function incorporated into the
operation of an atomic clock can never be unconditionally known. The
language of quantum theory reflects within itself a distinction between
`explanation' and `evidence'.  For explanations it offers the linear
algebra of wave functions and operators, while for evidence it offers
probabilities on a set of outcomes.  Outcomes are subject to quantum
uncertainty, but uncertainty is only the tip of an iceberg: how can one
``know'' that a wave function describes an experimental situation?  The
distinction within quantum theory between linear operators and
probabilities implies a gap between any explanation and the evidence
explained. \cite{ams02,aop05,tyler07,CUP}:
\begin{prop} \label{prop:one} To choose a wave function to explain 
experimental evidence requires reaching beyond logic based on that
evidence, and evidence acquired after the choice is made can call for a
revision of the chosen wave function.
\end{prop} 
Because no wave function can be unconditionally known, not even
probabilities of future evidence can be unconditionally foreseen.
Here we show implications of the unknowability of wave functions for
the second as a unit of measurement in the International System (SI),
implications that carry over to both digital communications and to the
use of a spacetime with a metric tensor in explaining clock readings
at the transmission and reception of logical symbols.

Clocks that generate Universal Coordinated Time (UTC)
are steered toward aiming points that depend not only on a chosen wave
function but also on an hypothesized metric tensor field of a curved
spacetime.  Like the chosen wave function, the hypothesis of a metric
tensor is constrained, but not determined, by measured data.
Guesses enter the operations of clocks through the computational machinery
that steers them.  Taking incoming data, the machinery updates records that
determine an aiming point, and so involves the writing and reading of
records.  The writing must take place at a phase of a cycle distinct from a
phase of reading, with a separation between the writing and the reading
needed to avoid a logical short circuit.  In Sec.~\ref{sec:turing} we
picture an explanation used in the operation of a clock as a string of
characters written on a tape divided into squares, one symbol per square.
The tape is part of a Turing machine modified to be stepped by a clock and
to communicate with other such machines and with keyboards and displays.
We call this modified Turing machine an \emph{open machine}. The
computations performed by an open machine are open to an inflow numbers and
formulas incalculable prior to their entry.  Because a computer cycles
through distinct phases of memory use, the most direct propagation of
symbols from one computer to another requires a symbol from one computer to
arrive during a suitable phase of the receiving computer's cycle.  In
Sec.~\ref{sec:phasing} we elevate this phase dependence to a principle that
defines the \emph{logical synchronization} necessary to a \emph{channel}
that connects clock readings at transmission of symbols to clock readings
at their reception

Recognizing the dependence of logic-bearing channels on an interaction
between evidence and hypotheses about signal propagation engenders
several types of questions, leading to a \emph{discipline of logical
  synchronization}, outlined in Sec.~\ref{sec:patterns}.
The first type of question concerns patterns of channels that are
possible aiming points, as determined in a blackboard calculation that
assumes a theory of signal propagation.  Sec.~\ref{sec:typeI}
addresses examples of constraints on patterns of channels under
various hypotheses of spacetime curvature, leading to putting ``phase
stripes'' in spacetime that constrain channels to or from a given open
machine.  An example of a freedom to guess an explanation within a
constraint of evidence is characterized by a subgroup of a group of
clock adjustments, and a bound on bit rate is shown to be imposed by
variability in spacetime curvature.

Sec.~\ref{sec:adj} briefly addresses the two other Types of questions,
pertaining not to \emph{hypothesizing} possible aiming points `on the
blackboard', but to \emph{using} hypothesized aiming points, copied
into feedback-mediating computers, for the steering of drifting
clocks. After discussing steering toward aiming points
copied from the blackboard, we note occasions that invite revision of
a hypothesized metric tensor and of patterns of channels chosen as
aiming points.

\section{Open Turing machine models a computer in a feedback loop}\label{sec:turing}
Computer-mediated feedback, especially as used in an atomic clock,
requires logic open to an inflow of inputs beyond the reach of
calculation.  To model the logic of a computer that communicates with
the other devices in a feedback loop, we modify a Turing machine to
communicate with external devices, including other such machines.  The
Turing machine makes a record on a tape marked into squares, each
square holding one character of an alphabet.  Operating in a sequence
of `moments' interspersed by `moves', at any moment the machine scans
one square of the tape, from which it can read, or onto which it can
write, a single character. A move as defined in the mathematics of
Turing machines consists (only) of the logical relation between the
machine at one moment and the machine at the next moment
\cite{turing}, thus expressing the logic of a computation, detached
from its speed; however, in a feedback loop, computational speed matters. Let
the moves of the modified Turing machine be stepped by ticks of a
clock.  A step occurs once per a period of revolution of the clock
hand.  This period is adjustable, on the fly.  We require that the
cycle of the modified Turing machine correspond to a unit interval of
the readings of its clock.

To express communication between open machines as models of computers,
the modified Turing machine can receive externally
supplied signals and can transmit signals, with both the reception and
the transmission geared to the cycle of the machine.  In addition, the
modified Turing machine registers a count of moments at which signals
are received and moments at which signals are transmitted.  At a finer
scale, \textit{the machine records a phase quantity in the cycle of
  its clock, relative to the center of the moment at which a signal
  carrying a character arrives.}  We call such a machine an \emph{open
  machine}.  An open machine can receive detections and can command
action, for instance the action of increasing or decreasing the
frequency of the variable oscillator of an atomic clock.

Calculations performed on an open machine
communicating with detectors and actuators proceed by moves made
according to a rule that can be modified from outside the machine in
the course of operation.  These calculations respond to received
influences, such as occurrences of outcomes underivable from the
contents of the machine memory, when the open machine writes
commands on a tape read by an external actuator.  The wider physical
world shows up in an open machine as both (1) unforeseeable messages
from external devices and (2) commands to external devices.

We picture a real-time computer in a feedback loop
as writing records on the tape of an open machine.  The segmentation into
moments interspersed by moves is found not just in Turing machines but
in any digital computer, which implies
\begin{prop}
The logical result of any computation is oblivious to variations in
speed at which the clock steps the computer\vspace*{6pt}.
 \end{prop}\\
\textsc{Corollary 2.1}. \textit{No computer can sense directly any variation in
its clock frequency.}
\vspace*{6pt}

Although it cannot directly sense variation in the tick rate of its
clock, the logic of open machine stepped by an atomic clock can still
control the adjustment of the clock's oscillator by responding to
variations in the detection rate written moment by moment onto its
Turing tape.  A flow of unforeseeable detections feeds successive
computations of results, each of which, promptly acted on, impacts
probabilities of subsequent occurrences of outcomes, even though those
subsequent outcomes remain unforeseeable.  The computation that steers
the oscillator depends not just on unforeseeable inputs, but also on a
steering formula encoded in a program. \vspace*{6pt}

\noindent\textbf{Remarks}:
\begin{enumerate}
\item To appreciate feedback, take note that a formula is distinct
  from what it expresses.  For example a formula written along a
  stretch of a Turing tape as a string of characters can contain a
  name $\psi$ for wave function as a function of time variable $t$ and
  space variables.  The formula, containing $\psi$, once written, just
  ``sits motionless,'' in contrast to the motion that the formula
  expresses.
\item Although unchanged over some cycles of a feedback loop, a
  feedback loop operates in a larger context, in which steering
  formulas are subject to evolution.  Sooner or later, the string that
  defines the action of an algorithm, invoking a formula, is apt to be
  overwritten by a string of characters expressing a new formula.
  Occasions for rewriting steering formulas are routine in clock
  networks, including those employed in geodesy and astronomy.
\end{enumerate}

\section{Communication channels and logical synchronization}\label{sec:phasing} 
Logical communication requires clocking.  The reading of a
clock of an open machine $A$---an $A$-reading---has the form
$m.\phi_m$ where $m$ indicates the count of cycles and $\phi_m$ is the
phase within the cycle, with the convention that $-1/2<\phi_m\le
1/2$.  We define a channel from $A$ to $B$, denoted $\abr$, as a set
of pairs, each pair of the form $(m.\phi_m,n.\phi_n)$.  The first
member $m.\phi_m$ is an $A$-reading at which machine $A$ can transmit
a signal and $n.\phi_n$ is a $B$-reading at which the clock of machine
$B$ can register the reception of the signal.  Define a
\emph{repeating channel} to be a channel $\abr$ such that
\begin{equation}
  (\forall \ell \in 
[\ell_1,\ell_2])(\exists m,n,j, k) (m+\ell j.\phi_{A,\ell},n+\ell k.\phi_{B,\ell})]
\in \abr,
\end{equation}
For theoretical purposes, it is convenient to define an
\emph{endlessly repeating channel} for which $\ell$ ranges over all integers.
Again for theoretical purposes, on occasion we consider channels
for which the phases are all zero, in which case we may omit writing the
phases.

Because they are defined by local clocks without reference to any
metric tensor, channels invoke no assumption about a metric or even a
spacetime manifold.  For this reason evidence from the operation of
channels is independent of any explanatory assumptions involving a
manifold with metric and, in particular, is independent of any global
time coordinate, or any ``reference system'' \cite{soffel03}. Thus
clock readings at the transmission and the reception of signals can
prompt revisions of hypotheses about a metric tensor field.  A record
format for such evidence was illustrated in earlier work
\cite{spie2011,1639}, along with the picturing of such records as
\emph{occurrence graphs}.

From the beating of a heart to the bucket brigade, life moves in
phased rhythms.  For a symbol carried by a signal from an open machine
$A$ to be written into the memory of an open machine $B$, the signal
must be available at $B$ within a phase of the cycle of $B$ during
which writing can take place, and the cycle must offer room for a
distinct other phase.  We elevate engineering commonplace to a
principle pertaining to open machines as follows.
\begin{prop}\label{prop:three} A logical symbol can propagate from one
open machine to another only if the symbol arrives within the writing
phase of the receiving machine; in particular, respect for phasing
requires that for some positive $\eta$ any arrival phase $\phi_n$
satisfy the inequality 
\begin{equation}\label{eq:main} |\phi_n| < (1-\eta)/2.  
\end{equation} 
\end{prop} 
Prop. \ref{prop:three} serves as a fixed point to hold onto while
hypotheses about signal propagation in relation to channels are
subject to revision.  We call the phase constraint on a channel
asserted by (\ref{eq:main}) \emph{logical synchronization}.  For
simplicity and to allow comparing conditions for phasing with
conditions for Einstein synchronization, we take the engineering
liberty of allowing transmission to occur at the same phase as
reception, so that both occur during a phase interval satisfying
(\ref{eq:main}).  The alternative of demanding reception near values
of $\phi=1/2$ can be carried out with little extra
difficulty.  \vspace*{6pt}\\
\noindent\textbf{Remarks:}
\begin{enumerate}
\item Note that $\phi_n$ in the proposition is a phase of a cycle of a
  variable-rate clock that is \emph{not} assumed to be in any fixed
  relation to a proper clock as conceived in general relativity.
  Indeed, satisfying (\ref{eq:main}) usually requires the operation of
  clocks at variable rates.
\item The engineering of communications between computers commonly
  detaches the timing of a computer's receiver from that of the
  computer by buffering: after a reception, the receiver writes
  into a buffer that is later read by the computer\cite{meyr}.  In
  analyzing open machines we do without buffering, confining ourselves
  to character-by-character phase meshing as asserted in
  Prop. \ref{prop:three}, which offers the most direct communication
  possible.
\end{enumerate}

\section{A discipline of logical synchronization}\label{sec:patterns}
Given the definition of a channel and the condition (\ref{eq:main})
essential to the communication of logical symbols, three types of
questions \vspace*{4pt} arise:

\noindent\textbf{Type I:} What patterns of interrelated channels does
one try for as aiming \vspace*{4pt} points?

\noindent\textbf{Type II:} How can the steering of open machines be arranged to
  approach given aiming points within acceptable phase \vspace*{4pt} tolerances?

\noindent\textbf{Type III:} How to respond to deviations from aiming points beyond
\vspace{4pt}  tolerances?

\noindent Such questions point the way to exploring what might be called a
\emph{discipline of logical synchronization}. So far we
notice two promising areas of application within this discipline:
\begin{enumerate}
\item Provide a theoretical basis for networks of logically synchronized
  repeating channels, highlighting 
  \begin{enumerate}
  \item possibilities for channels with null receptive phases as a
    limiting case of desirable behavior, and
\item circumstances that force non-null phases.
  \end{enumerate}
\item Explore constraints on receptive phases imposed by
  gravitation, as a path to exploring and measuring gravitational
  curvature, including slower changes in curvature than those searched
  for by the Laser Gravitational Wave Observatory \cite{ligo}.
\end{enumerate}

\subsection{Geometry of signal propagation}
Answers to questions of the above Types require hypotheses, if only
provisional, about signal propagation.  For this section we assume that
propagation is described by null geodesics in a Lorentzian 4-manifold $M$
with one or another metric tensor field $g$, as in general relativity.
Following Perlick \cite{perlick} we represent an open machine as a timelike
worldline, meaning a smooth embedding $\gamma\colon \zeta \mapsto
\gamma(\zeta)$ from a real interval into $M$, such that the tangent vector
$\dot{\gamma}(\zeta)$ is everywhere timelike with respect to $g$ and
future-pointing.  We limit our attention to worldlines of open machines
that allow for signal propagation between them to be expressed by null
geodesics.  To say this more carefully, we distinguish the \emph{image} of
a worldline as a submanifold of $M$ from the worldline as a mapping.
Consider an open region $V$ of $M$ containing a smaller open region $U$,
with $V$ containing the images of two open machines $A$ and $B$, with the
property that every point $a$ of the image of $A$ restricted to $U$ is
reached uniquely by one future-pointing null geodesic from the image of $B$
in $V$ and by one past-pointing null geodesic from the image of $B$ in
$V$. We then say $A$ and $B$ are \emph{radar linkable} in $U$.  We limit
our attention to open machines that are radar linkable in some spacetime
region $U$.  In addition we assume that the channels preserve order (what
is transmitted later arrives later).  Indeed, we mostly deal with open
machines in a gently curved spacetime region, adequately described by Fermi
normal coordinates around a timelike geodesic.

For simplicity and to allow comparing conditions for phasing with
conditions for Einstein synchronization, we take the liberty of allowing
transmission to occur at the same phase as reception, so that both occur
during a phase interval satisfying (\ref{eq:main}).  The perhaps more
realistic alternative of demanding reception near values of $\phi=1/2$ can
be carried out with little difficulty.

To develop the physics of channels, we need to introduce three
concepts:

(1) We define a \emph{group of clock adjustments} as transformations of the
readings of the clock of an open machine.  As it pertains to endlessly
repeating channels, a group $H$ of clock adjustments consists of functions
on the real numbers having continuous, positive first derivatives.  Group
multiplication is the composition of such functions, which, being
invertible, have inverses.  To define the action of $H$ on clock readings,
we speak `original clock readings' as distinct from 'adjusted readings' An
adjustment $f_A \in H$ acts by changing every original reading $\zeta_A$ of
a clock $A$ to an adjusted reading $f_A(\zeta_A)$.  As we shall see, clock
adjustments can affect echo counts.

(2) To hypothesize a relation between the $A$-clock and an
accompanying proper clock, one has to assume one or another metric
tensor field $g$, relative to which to define proper time increments
along $A$'s worldline; then one can posit an adjustment $f_A$ such that
$f_A(\zeta_A)=\tau_A$ where $\tau_A$ is the reading imagined for the
accompanying proper clock when $A$ reads $\zeta_A$.

(3) We need to speak of positional relations between open
machines. For this section we assume that when an open machine $B$
receives a signal from any other machine $A$ then $B$ echoes back a
signal to $A$ right away, so the echo count $\Delta_{ABA}$ defined
in Sec.~\ref{sec:phasing} involves no delay at $B$.  In this case,
evidence in the form of an echo count becomes explained, under the
assumption of a metric tensor field $g$, as being just twice the radar
distance \cite{perlick} from $A$ to the event of reception by $B$.

\section{Type-I questions: mathematical expression of possible 
patterns of channels}\label{sec:typeI} Questions of Type I concern
constraints on channels imposed by the physics of signal propagation.  Here
we specialize to constraints on channels imposed by spacetime metrics,
constraints obtained from mathematical models that, while worked out so to
speak on the blackboard, can be copied onto Turing tapes as aiming points
toward which to steer the behavior of the clocks of open machines.
Questions of Types II and III are deferred to the Sec. \ref{sec:adj}.

\subsection{Channels with null phases as aiming points: two open
machines linked by a two-way channel} We begin by considering just two
machines.  Assuming an hypothetical spacetime $(M,g)$, suppose that machine
$A$ is given as a worldline parametrized by its clock readings: what are
the possibilities and constraints for an additional machine $B$ with
two-way repeating channels $\abr$ and $\bar$ with a constant
echo count? We assume the idealized case of channels with null phases,
which implies integer echo counts.  For each $A$-tick there is a future
light cone and a past light cone.
\begin{figure}[h]
\centerline{\includegraphics[width=4.9in]{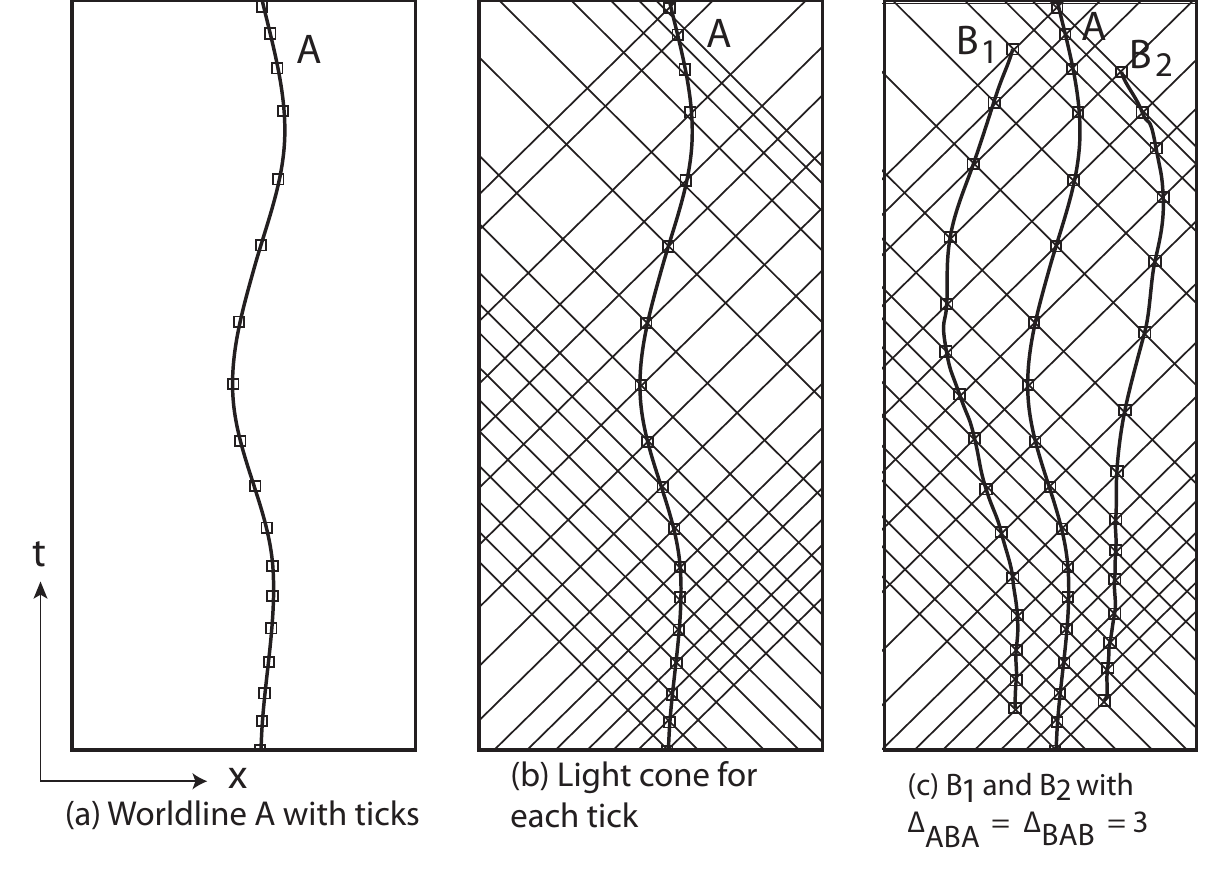}}
\caption{(a) Worldline of $A$ with tick events indicated; (b) Light
  cones associated to ticks of $A$; (c) Ticks of $B_1$ and $B_2$ at light cone
  intersections corresponding to $\Delta_{ABA}=\Delta_{BAB}=3$.
}\label{fig:3}
\end{figure}
The future light cone from an $A$-reading
$\zeta_A=m$ has an intersection with the past light cone for the
returned echo received at $\zeta_A=m+\Delta_{ABA}$.  Fig.~\ref{fig:3}
illustrates the toy case of a single space dimension in a flat
spacetime by showing the two possibilities for a machine $B$ linked to
$A$ by two-way channels at a given constant echo count.  In each
solution, the clocking of $B$ is such that a tick of $B$ occurs at
each of a sequence of intersections of outgoing and incoming light
cones from and to ticks of $A$.  Note that the image of $B$, and not
just its clock rate, depend on the clock rate of $A$.

Determination of the tick events for $B$ leaves undetermined the $B$
trajectory between ticks, so there is a freedom of choice.  One can
exercise this freedom by requiring the image of $B$ to be consistent
with additional channels of larger echo counts. A clock adjustment of
$A$ of the form $\zeta_A\rightarrow \zeta'_A=N\zeta_A$ for $N$ a
positive integer increases the density of the two-way channel by $N$
and inserts $N-1$ events between successive $B$-ticks, thus
multiplying the echo count by $N$.  As $N$ increases without limit,
$B$ becomes fully specified.

Turning to two space dimensions, the image of $B$ must lie in a tube
around the image of $A$, as viewed in a three-dimensional space
(vertical is time).  So any timelike trajectory within the tube will
do for the image of $B$.  For a full spacetime of 3+1 dimensions, the
solutions for the image of $B$ fall in the corresponding
``hypertube.'' The argument does not depend on flatness and so works
for a generic, gently curved spacetime in which the channels have the
property of order preservation.

\begin{figure}[h]
\centerline{\includegraphics[width=4.9in]{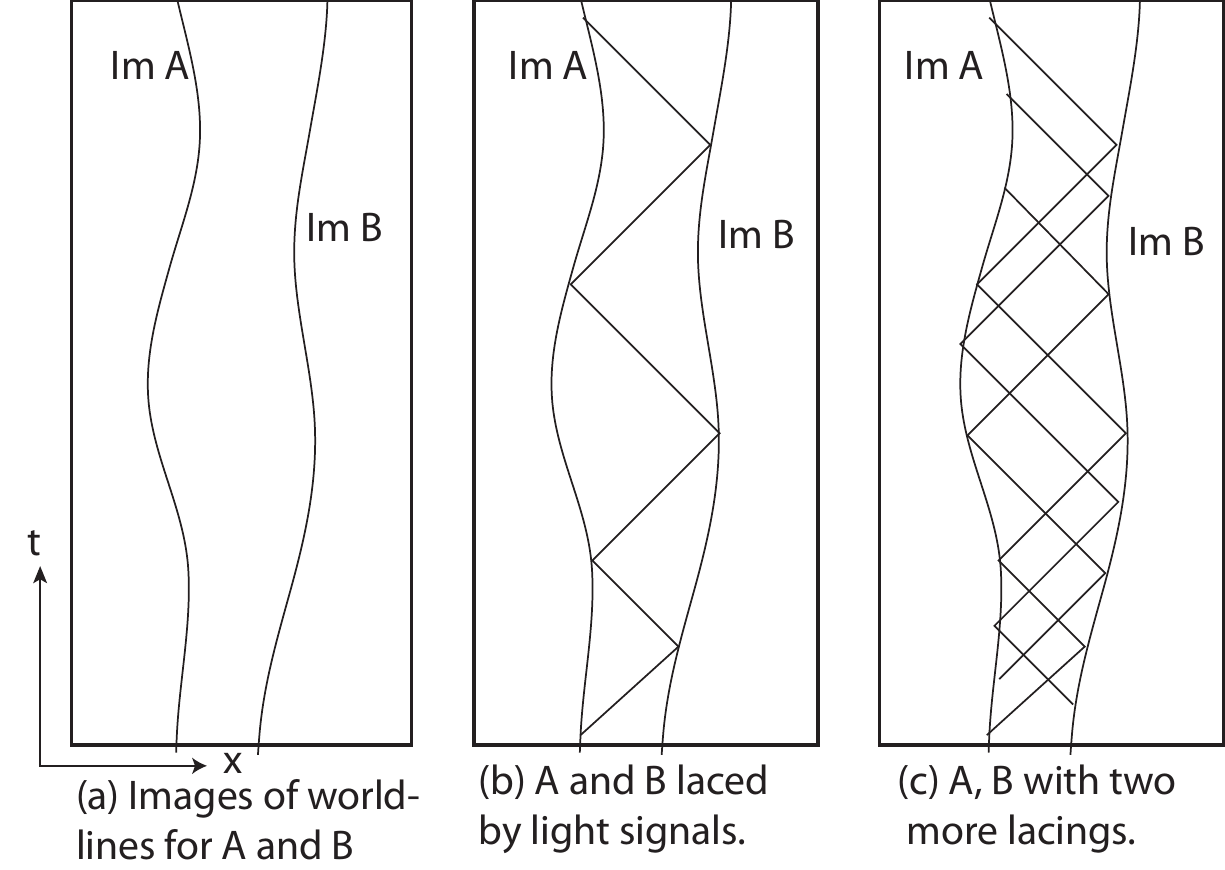}}
\caption{(a) Images of worldlines for open machines $A$ and $B$ freely
  chosen; (b) Light signals lacing $A$ and $B$ define tick events; (c)
  Interpolated lacings of light signals added to make
  $\Delta_{ABA}=\Delta_{BAB}=3$.}\label{fig:4}
\end{figure}

A different situation for two machines arises in case only the image
of $A$'s worldline is specified while its clocking left to be
determined.  In this case the image of $B$ can be freely chosen, after
which the clocking of both $A$ and $B$ is constrained, as illustrated
in Fig.~\ref{fig:4} for the toy case of flat spacetime with 1 space
dimension.  To illustrate the constraint on clocking, we define a
``lacing'' of light signals to be a pattern of light signals echoing
back and forth between two open machines as illustrated in
Fig.~\ref{fig:4}~(b).  For any event chosen in the image of $A$, there
is a lacing that touches it.  In addition to choosing this event, one
can choose any positive integer $N$ to be $\Delta_{ABA}$, and choose
$N-1$ events in the image of $A$ located after the chosen event and
before the next $A$-event touched by the lacing of light signals.  The
addition of lacings that touch each of the $N-1$ intermediating events
corresponds to a repeating channel $\abr$ with echo count
$\Delta_{ABA}=N$, along with a repeating channel $\bar$ with the same
echo count $\Delta_{BAB}=N$.  This construction does not depend on the
dimension of the spacetime nor on its flatness, and so works also for
a curved spacetime having the property of order preservation.

\subsection{Example of free choice characterized by a transformation group}\label{sec:group}
Evidence of channels as patterns of clock readings leaves open a choice of
worldlines for its explanation.  In the preceding example of laced channels
between open machines $A$ and $B$, part of this openness can be reflected
within analysis by the invariance of the channels under a subgroup of the
group of clock adjustments that ``slides the lacings,'' as follows.
Suppose that transmissions of an open machine $A$ occur at given values of
$A$-readings.  We ask about clock adjustments that can change the events of
a worldline that correspond to a given $A$-reading.  If a clock adjustment
$f_A$ takes original $A$-readings $\zeta_A$ to a revised $A$-readings
$f_A(\zeta_A)$, transmission events triggered by the original clock
readings become triggered when the re-adjusted clock exhibits the
\emph{same readings}.  As registered by original readings, the adjusted
transmission occurs at $\zeta'_A=f_A^{-1}(\zeta_A)$.  Based on this
relation we inquire into the action of subgroups of $H\times H$ on the
readings of the clocks of two open machines $A$ and $B$.  In particular,
there is a subgroup $K(A,B) \subset H\times H$ that expresses possible
revisions of explanations that leave invariant the repeating channels with
constant echo count $N$.  An element $f_A\times f_B \in K(A,B)$ is a pair
of clock adjustments that leaves the channels invariant, and such a pair
can be chosen within a certain freedom.  For the adjustment $f_A$ one is
free to: (a) assign an arbitrary value to $f_A^{-1}(0)$; and (b), if $N>1$,
then for $j,k=1,\ldots,N-1$, choose the value of $f_A^{-1}(j)$ at will,
subject to the constraints that $k>j\Rightarrow f^{-1}(k)>f^{-1}(j)$ and
$f^{-1}(N-1)$ is less than the original clock reading for the re-adjusted
first echo from $f^{-1}(0)$.  With these choices, $f_B$ is then constrained
so that each lacing maps to another lacing.  The condition (a) slides a
lacing along the pair of machines; the condition (b) nudges additional
lacings that show up in the interval between a transmission and the receipt
of its echo.  In this way a freedom to guess within a constraint is
expressed by $K(A,B$.

\subsection{Channels among more than two open machines}
Moving to more than two machines, we invoke the
\begin{quote}
  \textbf{Definition:} an \emph{arrangement of open machines} consists of
  open machines with the specification of some or all of the channels from
  one to another, augmented by proper periods of the clock of at least one
  of the machines.
\end{quote}
(Without specifying some proper periods, the scale of separations of
one machine from another is open, allowing the arrangement to shrink
without limit, thus obscuring the effect of spacetime curvature.)

Although gentle spacetime curvature has no effect on the possible
channels linking two open machines, curvature does affect
the possible channels and their echo counts in some arrangements of
five or more machines, so that the possible arrangements
are a measure spacetime curvature.
\begin{figure}[h]
\centerline{\includegraphics[width=4.9in]{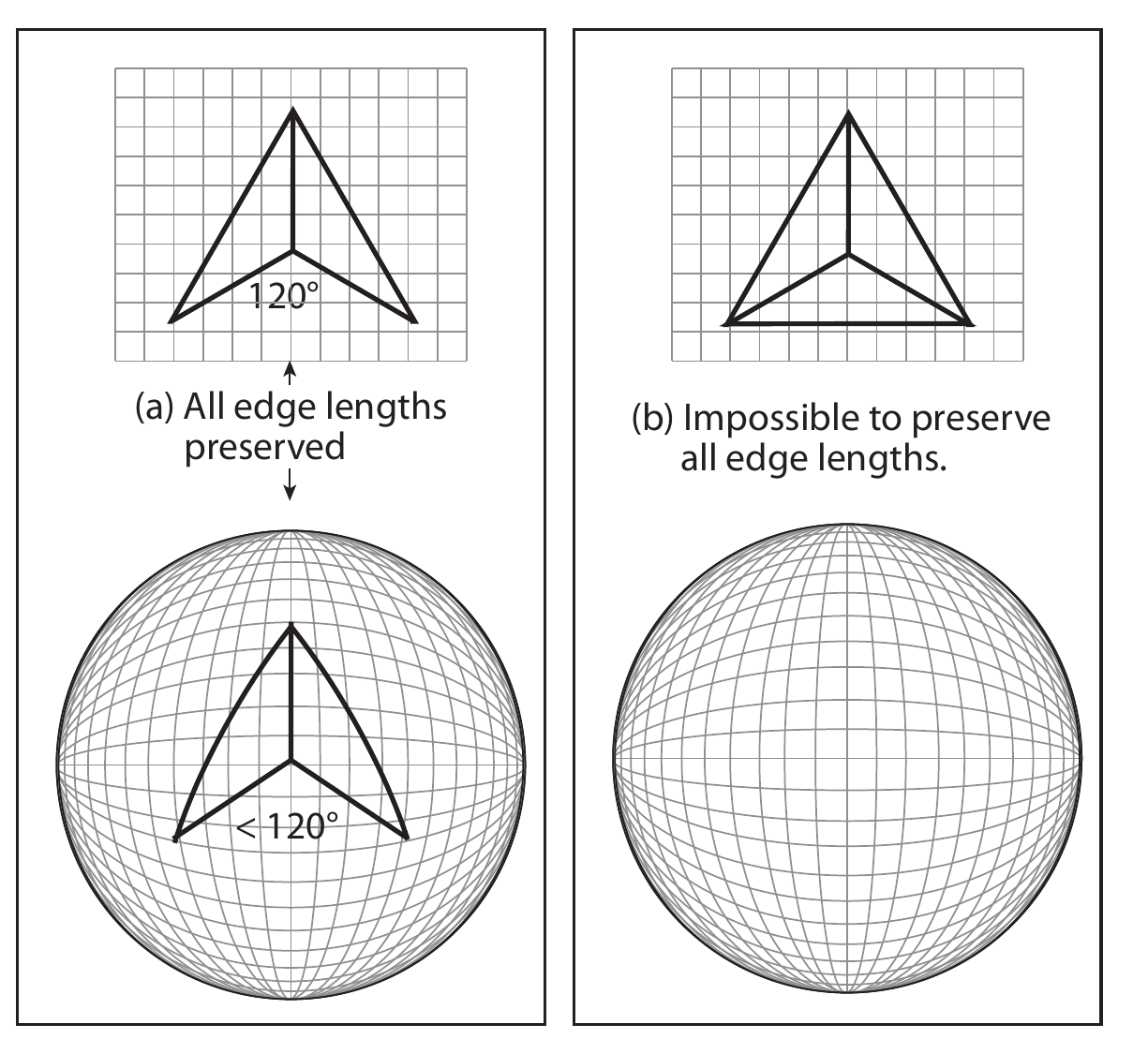}}
\caption{Plane figures, one of which maps to a sphere while preserving edge lengths}
\label{fig:sphere}
\end{figure}
The way that spacetime curvature affects the possible arrangements of
channels is analogous to the way surface curvature in Euclidean geometry
affects the ratios of the lengths of the edges of embedded graphs.  The
effect on ratios shows up in mappings from graphs embedded in a plane to
their images on a sphere.  For example, a triangle can be mapped from a
plane to a generic sphere, in such a way that each edge of the triangle is
mapped to an arc of the same length along a great circle on the sphere.
The same holds for two triangles that share an edge, as illustrated in
Fig.~\ref{fig:sphere}, panel (a); however, the Gauss curvature of the
sphere implies that the complete graph on 4 vertices generically embedded
in the plane, shown in panel (b), cannot be mapped so as to preserve all
edge lengths.  The property that blocks the preservation of edge ratios is
the presence of an edge in the plane figure that cannot be slightly changed
without changing the length of at least one other edge; we speak of such an
edge as ``frozen.''

In a static spacetime, which is all we have so far investigated, a generic
arrangement of 4 open machines, is analogous to the triangle on the plane
in that a map to any gently curved spacetime can preserve all the echo
counts.
\begin{prop}\label{prop:nine}
Assume four open machines in a static spacetime, with one machine
stepped with a proper-time period $p_\tau$, and let $N$ be any
positive integer.  Then, independent of any gentle Riemann curvature
of the spacetime, the four open machines can be arranged, like
vertices of a regular tetrahedron, to have six two-way channels with
null phases, with all echo counts being $2N$.
\end{prop}
\textit{Proof:} Assuming a static spacetime, choose a coordinate system
with all the metric tensor components independent of the time coordinate,
in such a way that it makes sense to speak of a time coordinate distinct
from space coordinates (for example, in a suitable region of a
Schwarzschild geometry).  Let$V_1$ denote the machine with specified proper
period $p_{\tau}$, and let $V_2$, $V_3$, and $V_4$ denote the other three
machines.  For $i,j \in \{1,2,3,4\}$, $i \ne j$, we prove the possibility,
independent of curvature, of  the channels
\begin{equation}\label{eq:Vs}
\stackrel{\xrightarrow{\hspace*{0.8cm}}}{V_iV_j}=\{(k,k+N.0)|k \text{
  any integer}\}.  
\end{equation}
Require that each of four machines be located at some fixed spatial
coordinate.  Because the spacetime is static, the coordinate time
difference between a transmission at $V_1$ and a reception at any other
vertex $V_j$ (a) is independent of the value of the time coordinate at
transmission and (b) is the same as the coordinate time difference between
a transmission at $V_j$ and a reception at $V_1$.  For this reason any
one-way repeating channel of the form (\ref{eq:Vs}) can be turned around to
make a channel in the opposite direction, so that establishing a channel in
one direction suffices.  For transmissions from any vertex to any other
vertex, the coordinate-time difference between events of transmission
equals the coordinate time difference between receptions. A signal from a
transmission event on $V_1$ propagates on an expanding light cone, while an
echo propagates on a light cone contracting toward an event of reception on
$V_1$.  Under the constraint that the echo count is $2N$, (so the proper
duration from the transmission event to the reception event for the echo is
$2N p_{\tau}$), the echo event must be on a 2-dimensional submanifold---a
sphere, defined by constant radar distance $N p_{\tau}$ of its points
from $V_1$ with transmission at a particular (but arbitrary) tick of $V_1$.
In coordinates adapted to a static spacetime, this sphere may appear as
a ``potatoid'' in the space coordinates, with different points on the
potatoid possibly varying in their time coordinate.  The potatoid shape
corresponding to an echo count of $2N$ remains constant under evolution of
the time coordinate.  Channels from $V_1$ to the other three vertices
involve putting the three vertices on this potatoid.  Put $V_2$
anywhere on the ``potatoid''.  Put $V_3$ anywhere on the ring that is
intersection of potatoid of echo count $2N$ radiated from $V_2$ and that
radiated from $V_1$.  Put $V_4$ on an intersection of the potatoids
radiating from the other three vertices.\\ Q.E.D.

According to Prop \ref{prop:nine} the channels, and in particular the
echo counts possible for a complete graph of four open machines in
flat spacetime are also possible for a spacetime of gentle static
curvature, provided that three of the machines are allowed to set
their periods not to a fixed proper duration but in such a way that
all four machines have periods that are identical in coordinate time.
The same holds if fewer channels among the four machines are
specified.

But for five machines, the number of channels connecting them matters.
Five open machines fixed to space coordinates in a static spacetime are
analogous to the 4 vertices of a plane figure, in that an arrangement
corresponding to an incomplete graph on five vertices can have echo counts
independent of curvature, while a generic arrangement corresponding to a
complete graph must have curvature-dependent relations among its echo
counts.

\begin{prop}\label{prop:9.5}
Assuming a static spacetime, consider an arrangement of five open
machines obtained by starting with a tetrahedral arrangement of four
open machines with all echo counts of $2N$ as in
Prop. \ref{prop:nine}, and then adding a fifth machine: independent of
curvature, a fifth open machine can be located with two-way channels
having echo counts of $2N$ linking it to any three of the four
machines of tetrahedral arrangement, resulting in nine two-way
channels altogether.
\end{prop}
\textit{Proof:} The fifth machine can be located as was the machine
$V_4$, but on the side opposite to the cluster $V_1$, $V_2$, $V_3$.\\
Q.E.D.

\begin{figure}[h]
\centerline{\includegraphics[width=4.9 in]{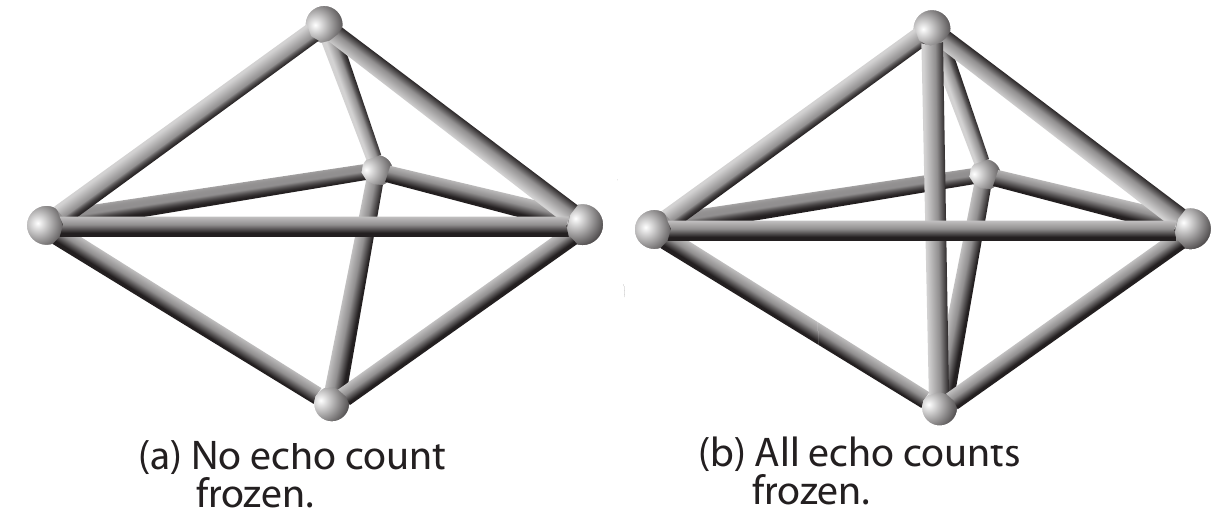}}
\caption{(a) 5 open machines with 9 two-way channels; (b) 
Five open machines with all 10 two-way channels}\label{fig:5pt}
\end{figure}

In contrast to the arrangement of 9 two-way channels, illustrated in
Fig.~\ref{fig:5pt} (a) consider an arrangement of 5 open machines
corresponding to a complete graph on five vertices, with has ten two-way
channels, as illustrated in Fig.~\ref{fig:5pt} (b).  For five open machines
in a generic spacetime, not all of the ten two-way channels
can have the same echo counts.  Instead, channels in a flat spacetime as
specified below can exist with about the simplest possible ratios of echo
counts.  Label five open machines, $A_1$, $A_2$, $A_3$, $B_1$, and $B_2$.
Take $B_1$ to be stepped by a clock ticking at a fixed proper period
$p_\tau$, letting the other machines tick at variable rates to be
determined.  Let $X$ be any machine other than $B_1$.  For a flat spacetime
it is consistent for the proper periods of all 5 machines to be $p_\tau$,
for the echo counts $\Delta_{B_1XB_1}$ to be $4N$ and for the echo counts
$\Delta_{A_i A_jA_i}$ to be $6N$, leading to twenty channels, conveniently
viewed as in Fig.~\ref{fig:5pt} (b) as consisting of ten two-way channels.
\begin{prop}\label{prop:ten}
Consider 5 open machines each fixed to space coordinates in a static curved
spacetime in which the machines are all pairwise radar linkable, with
10 two-way channels connecting each machine to all the others; then:
\begin{enumerate}
\item Allowing for the periods of the machines other than $B_1$ to vary, it
  is consistent with the curvature for all but one of the ten two-way
  channels to have null phases and echo counts as in a flat spacetime, but
  at least one two-way channel must have a different echo count that
  depends on the spacetime curvature.
\item Suppose $m$ of the 10 two-way links are allowed to have non-zero
  phases.  If the spacetime does not admit all phases to be null, in
  generic cases the least possible maximum amplitude of a phase
  decreases as $m$ increases from 1 up to 10.
\item The periods of the clocks of the open machines can be taken to
  be the coordinate-time interval corresponding to the proper period
  $p_\tau$ at $B_1$.
\end{enumerate}
\end{prop}
\textit{Proof:} Reasoning as in the proof of Prop.\  \ref{prop:nine} with
its reference to a static spacetime shows that the same echo counts
are possible as for flat spacetime \emph{with the exception} that at
least one of the two-way channels must be free to have a different
echo count.  For $m < 10$, similar reasoning shows that allowing $m+1$
machines vary in echo count allows reduction in the maximum variation
from the echo counts in a flat spacetime, compared to the case in
which only $m$ machines are allowed to vary in echo count.\\Q.E.D.

Adding the tenth two-way channel to an arrangement of five open
machines effectively ``freezes'' all the echo counts.  To define
``freezing'' as applied to echo counts, first take note an asymmetry
in the dependence of echo counts on clock rates.  Consider any two
machines $A$ and $B$; unlike echo count $\Delta_{BAB}$, which $B$ can
change by changing it clock rate, the echo count $\Delta_{ABA}$ is
insensitive to $B$'s clock rate.  An echo count $\Delta_{ABA}$ will be
said to be \emph{to} $B$ and \emph{from} $A$.
\begin{quote}
  \textbf{Definition:} An arrangement of open machines is
  \emph{frozen} if it has an echo count to a machine $B$ that cannot
  be changed slightly without changing the length of another echo
  count to $B$.
\end{quote}
The property of being frozen is important because of the following.
\begin{prop}
  Whether or not a frozen arrangement of open machines is consistent
  with an hypothesized spacetime depends on the Weyl curvature of
  the spacetime.
\end{prop}

For example, think of the 5 open machines as carried by 5 space vehicles
coasting along a radial geodesic in a Schwarzschild geometry.  In this
example the variation of echo counts with curvature is small enough to be
expressed by non-null phases of reception.  In Fermi normal coordinates
centered midway between the radially moving open machine $B_1$ and $B_2$
one has the metric with a curvature parameter $\mu:=GM/(c^2r^3)$, where $r$
is the Schwarzschild radial coordinate to the origin of the Fermi normal
coordinates, $x$ is the radial distance coordinate from from the center
point between $B_1$ and $B_2$, and $y$ and $z$ are transverse to the radial
direction along which $B_j$ coasts\cite{manasse}.  The speed of light is
$c$.  We make the adiabatic approximation which ignores the time dependence
of $r$, so that in calculations to first order in curvature we take
advantage of the (adiabatically) static spacetime by locating open machines
at fixed values of $x,y,z$.  The metric is symmetric under rotation about
the (radially directed) $x$-axis.  Let $B_1$ and $B_2$ be located
symmetrically at positive and negative values, respectively, of the
$x$-axis, and let $A_0$, $A_1$, and $A_2$ be located on a circle in the
plane $x=0$.  With the five machines so located, the coordinate-time
difference between transmissions is then the same as the coordinate-time
difference between receptions, and the coordinate-time delay in one
direction equals that in the opposite direction (as stated in the proof of
Prop \ref{prop:nine}).  We construct seven two-way channels as above with
null phases and show that the remaining 3 two-way channels can have the
equal phases, but that this phase must be non-null with a curvature
dependent amplitude $\phi$.
\begin{prop}\label{prop:eleven}
Under the stated conditions, if $\mu p_\tau^2c^2$ is small enough so
that\\ $27N \mu N^2p_\tau^2c^2/8 < 1/2$, then $ \phi= -27GM
N^3p_\tau^2/(8r^3)$
\end{prop}

\subsection{Changing curvature limits bit rate}
For a fixed separation $L$ between $B_1$ and $B_2$, an adiabatic
change in curvature imposes a constraint on bit rate possible for the
channels, stemming from a lower bound on clock periods.
Suppose the cluster of 5 open machines is arranged so that the
proper radar distance $L$ from $B_1$ to $B_2$ is 6,000 km, suppose the
cluster descends from a great distance down to a radius of $r=30,000$ km
from an Earth-sized mass $M_{\oplus} = 6.67\times 10^{24}$ kg.  
For simplicity, assume that the positions and clock rates are
continually adjusted to maintain null phases for all but
the three channels
$\stackrel{\xrightarrow{\hspace*{0.9cm}}}{A_nA_{n\pm 1}}$.
Because  $L\approx 2Np_\tau c$,
Prop. \ref{prop:eleven} implies  $\phi= - 23ML^3/(48r^3c^3p_{\tau})$,
which with (\ref{eq:main}) implies that $
  p_{\tau}>27 G M L^3/(32r^3c^3)$.

Substituting the parameter values, one finds that for the phases for
the channels $\stackrel{\xrightarrow{\hspace*{0.9cm}}}{A_nA_{n\pm 1}}$
to satisfy (\ref{eq:main}), it is necessary that
$p_\tau> 1.0\cdot 10^{-13}$ s.
If an alphabet conveys $b$ bits/character, the maximum bit rate
for all the channels in the 5-machine cluster is
$b/p_\tau<10^{13}b$ bits/s.

\section{Steering while listening to the
  unforeseeable}\label{sec:adj}

Turning from Type I to questions of Type II, we look at how the preceding
``blackboard modeling'' of clocks, expressed in the mathematical language
of general relativity, get put to work when models are encoded into the
open machines that manage their own logical synchronization.  For questions
of Type II (and Type III) both models that explain or predict evidence and
the evidence itself, pertaining to physical clocks, come into play.  Models
encoded into computers contribute to the steering of physical clocks in
rate and relative position toward an aiming point, generating echo counts
as evidence that, one acquired, can stimulate the guessing of new models
that come closer to the aiming point.

To express the effect of quantum uncertainty on logical
synchronization, specifically on deviations from aiming points, one
has to bring quantum uncertainty into cooperation with the
representation of clocks by general-relativistic worldlines.  This
bringing together hinges on distinguishing evidence from its
explanations.  Timelike worldlines and null geodesics in explanations,
being mathematical, can have no \emph{mathematical} connection to
physical atomic clocks and physical signals. To make such a connection
them one has to invoke the logical freedom to make a guess.  Within
this freedom, one can resort to quantum theory to explain deviations
of an atomic clock from an imagined proper clock, represented as a
worldline, without logical conflict.  

\subsection{Need for prediction in steering toward an aiming point}
Because of quantum uncertainty and for other reasons, if an aiming
point in terms of channels and a given frequency scale is to be
reached, steering is required, in which evidence of deviations from
the aiming point combine with hypotheses concerning how to steer
\cite{medterm,algorithm}.  To keep things simple, consider a case of
an aiming point with null phases, involving two open machines $A$ and
$B$, as in the example of Sec.~\ref{sec:typeI}, modeled by a given
worldline $A$ with given clock readings $\zeta_A$, where $B$ aims to
maintain two-way, null-phase channel of given $\Delta_{ABA}=
\Delta_{BAB}$.  For this $B$ registers arriving phases of reception
and adjusts its clock rate and its position more or less continually
to keep those phases small.  Deviations in position that drive
position corrections show up not directly at $B$ but as phases
registered by $A$, so the steering of machine $B$ requires information
about receptive phases measured by $A$.  The knowledge of the
deviation in position of $B$ at $\zeta_B$ cannot arrive at $B$ until
its effect has shown up at $A$ and been echoed back as a report to
$B$, entailing a delay of at least $\Delta_{BAB}$, hence requiring
that machine $B$ predict the error that guides for $\Delta_{BAB}$
prior to receiving a report of the error.  That is, steering
deviations by one open machine are measured in part by their effect on
receptive phases of other open machines, so that steering of one
machine requires information about receptive phases measured by other
machines, and the deviations from an aiming point must increase with
increasing propagation delays that demand predicting further ahead.

As is clear from the cluster of five machines discussed in
Sec.~\ref{sec:typeI}, the aiming-point phases cannot in general all be
taken to be zero.  For any particular aiming-point phase $\phi_0$
there will be a deviation of a measured phase quantity $\phi$ given by
\begin{equation}
  \delta:= \phi-\phi_0
\end{equation}
Whatever the value of $\phi_0$, adjustments to contain phases within
tolerable bounds depends on phase changes happening only gradually, so
that trends can be detected and responded to on the basis of adequate
prediction (aka guesswork). \vspace{6pt}\\
\noindent\textbf{Remarks:}
\begin{enumerate}
\item Unlike cycle counts of open machines,
  which we assume are free of uncertainty, measured phases and
  deviations of phases from aiming points are quantities subject to
  uncertainty.  For logic to work in a network, transmission of
  logical symbols must preserve sharp distinctions among them; yet the
  maintenance of sharp distinctions among transmitted symbols requires
  responses to fuzzy measurements.
 \item The acquisition of logical synchrony in digital communications
   involves an unforeseeable waiting time, like the time for a coin on
   edge to fall one way or the other \cite{meyr}.
\end{enumerate}

\subsection{Adjusting the aiming point}\label{sec:adjAim} 
Aiming points are not forever, and here we say a few words about
questions of Type III, in which an aiming point based on a
hypothesized metric tensor appears unreachable, and perhaps needs to
be revised. We have so far looked at one or another manifold with
metric $(M,g)$ as some given hypothesis, whether explored on the
blackboard or coded into an open machine to serve in maintaining its
logical synchronization.  In this context we think of $(M,g)$ as
``given.''  But deviations of phases outside of tolerances present
another context, calling for revising a metric tensor field.  In this
context one recognizes that a metric tensor field is hypothesized
provisionally, to be revised as prompted by deviations outside allowed
tolerances in implementing an aiming point.

Drawing on measured phases as evidence in order to adjust a
hypothesis of a metric tensor is one way to view the operation of the
Laser Interferometer Gravitational-Wave Observatory (LIGO)
\cite{ligo}.  While LIGO sensitivity drops off severely below 45 Hz,
the arrangement of five open machines of Prop.~\ref{prop:ten} has no
low-frequency cutoff, and so has the potential to detect arbitrarily
slow changes in
curvature.


\begin{thebibliography}{99}
\bibitem{sp330} B. N. Taylor and A. Thompson, Eds, \textit{The
  International System of Units (SI)}, NIST Special Publication
  330, 2008 Edition, National Institutes of Science and
  Technology. 
\bibitem{ams02} J. M. Myers and F. H. Madjid, ``A proof that measured
  data and equations of quantum mechanics can be linked only by
  guesswork,'' in S. J. Lomonaco Jr. and H.E. Brandt (Eds.)
  \textit{Quantum Computation and Information}, Contemporary
  Mathematics Series, vol. 305, American Mathematical Society,
  Providence, 2002, pp. 221--244.
\bibitem{aop05} F.~H. Madjid and J.~M. Myers, 
``Matched detectors as definers of force,'' 
Ann.\ Physics \textbf{319}, 251--273 (2005).
\bibitem{tyler07} J.~M. Myers and F.~H. Madjid, ``Ambiguity in
  quantum-theoretical descriptions of experiments,'' in
  K. Mahdavi and D. Koslover, eds., \textit{Advances in Quantum
    Computation}, Contemporary Mathematics Series, vol.~482
  (American Mathematical Society, Providence, I, 2009),
  pp.\ 107--123.
\bibitem{CUP} J. M. Myers and F. H. Madjid, ``What probabilities
  tell about quantum systems, with application to entropy and
  entanglement,'' in A. Bokulich and G. Jaeger, eds., \textit{Quantum
    Information and Entanglement}, Cambridge University Press,
  Cambridge UK, pp. 127--150 (2010).
\bibitem{turing} A.~M. Turing, 
``On computable numbers with an application to the
Entscheidungsproblem,'' 
Proc.\ London Math.\ Soc., Series 2, \textbf{42}, 230--265 (1936--37).
\bibitem{soffel03} M. Soffel et al., ``The IAU resolutions
  for astrometry, celestial mechanics, and metrology in the
  relativistic framework: explanatory supplement,'' The
  Astronomical Journal, \textbf{126}, 2687--2706 (2003).
\bibitem{spie2011} J. M. Myers and F. H. Madjid, ``Rhythms essential
  to logical communication,'' in Quantum Information and Computation
  IX, E. Donkor, A. R. Pirich, and H. E. Brandt, eds, Proceedings of
  the SPIE,  \textbf{8057}, pp. 80570N1--12 (2011).  
\bibitem{1639} J. M. Myers and F. H. Madjid, ``Rhythms of Memory and
  Bits on Edge: Symbol Recognition as a Physical Phenomenon,''
  arXiv:1106.1639, 2011.
\bibitem{meyr} H. Meyr and G. Ascheid, \textit{Synchronization in
  Digital Communications}, Wiley, New York, 1990.
\bibitem{ligo} The LIGO Scientific Collaboration (http://www.ligo.org)
``LIGO: the Laser Interferometer Gravitational-Wave Observatory,''
Rep. Prog. Phys. \textbf{72}, 076901 (2009)
\bibitem{perlick} V. Perlick, ``On the radar method in
  general-relativistic spacetimes,'' in H. Dittus, C. L\"ammerzahl,
  and S. Turyshev, eds., \textit{Lasers, Clocks and Drag-Free
    Control: Expolation of Relativistic Gravity in Space}, (Springer,
  Berlin, 2008); also arXiv:0708.0170v1.
\bibitem{manasse} F. K. Manasse and C. W. Misner, J. Math Phys.,
  ``Fermi normal coordinates and some basic concepts in differential
  geometry,'' \textbf{4}, 735--745 (1963).
\bibitem{medterm} T. E. Parker, S. R. Jefferts, and T. P. Heavner,
  ``Medium-term frequency stability of hydrogen masers as measured by
  a cesium fountain,'' 2010 IEEE International Frequency Control
  Symposium (FCS), pp. 318--323 (2010). (available at
  http://tf.boulder.nist.gov/general/pdf/2467.pdf)
\bibitem{algorithm} J. Levine and T. Parker, ``The algorithm used to
  realize UTC(NIST),'' 2002 IEEE International Frequency Control
  Symposium and PDA Exhibition, pp. 537--542 (2002).
\end{thebibliography}
\end{document}